# From the Closed Classical Algorithmic Universe to an Open World of Algorithmic Constellations

Mark Burgin[1] and Gordana Dodig-Crnkovic[2]

[1] Dept. of Mathematics, UCLA, Los Angeles, USA. E-mail: mburgin@math.ucla.edu

[2] Mälardalen University, Department of Computer Science and Networks, School of Innovation, Design and Engineering, Västerås, Sweden; E-mail: gordana.dodig-crnkovic@mdh.se

**Abstract**

In this paper we analyze methodological and philosophical implications of algorithmic aspects of unconventional computation. At first, we describe how the classical algorithmic universe developed and analyze why it became closed in the conventional approach to computation. Then we explain how new models of algorithms turned the classical closed algorithmic universe into the open world of algorithmic constellations, allowing higher flexibility and expressive power, supporting constructivism and creativity in mathematical modeling. As Gödel's undecidability theorems demonstrate, the closed algorithmic universe restricts essential forms of mathematical cognition. In contrast, the open algorithmic universe, and even more the open world of algorithmic constellations, remove such restrictions and enable new, richer understanding of computation.

**Introduction**

Te development of various systems is characterized by a tension between forces of conservation (tradition) and change (innovation). Tradi-tion sustains system and its parts, while innovation moves it forward advancing some segments and weakening the others. Efficient functioning of a system depends on the equilibrium between tradition and innovation. When there is no equilibrium, system declines; too much tradition brings stagnation and often collapse under the pressure of inner or/and outer forces, while too much innovation leads to instability and frequently in rupture.

The same is true of the development of different areas and aspects of social systems, such as science and technology. In this article we are interested in computation, which has become increasingly important for society as the basic aspect of information technology. Tradition in computation is represented by conventional computation and classical algorithms, while unconventional computation stands for the far-reaching innovation.

It is possible to distinguish three areas in which computation can be unconventional:

1. *Novel hardware* (e.g. quantum systems) provides material realization for unconventional computation.

2. *Novel algorithms* (e.g. super-recursive algorithms) provide operational realization for unconventional computation.

3. *Novel organization* (e.g. evolutionary computation or self-optimizing computation) provides structural realization for unconventional computation.

Here we focus on algorithmic aspects of unconventional computation and analyze methodological and philosophical problems related to it,

making a distinction between three classes of algorithms: *recursive*, *subrecursive*, and *super-recursive algorithms*.

Each type of *recursive algorithms* form a class in which it is possible to compute exactly the same functions that are computable by Turing machines. Examples of recursive algorithms are partial recursive functions, RAM, von Neumann automata, Kolmogorov algorithms, and Minsky machines.

Each type of *subrecursive algorithms* forms a class that has less computational power than the class of all Turing machines. Examples of subrecursive algorithms are finite automata, primitive recursive functions and recursive functions.

Each type of *super-recursive algorithms* forms a class that has more computational power than the class of all Turing machines. Examples of super-recursive algorithms are inductive and limit Turing machines, limit partial recursive functions and limit recursive functions.

The main problem is that conventional types and models of algorithms make the algorithmic universe, i.e., the world of all existing and possible algorithms, closed because there is a rigid boundary in this universe formed by recursive algorithms, such as Turing machines, and described by the Church-Turing Thesis. This closed system has been overtly dominated by discouraging incompleteness results, such as Gödel incompleteness theorems.

Contrary to this, super-recursive algorithms controlling and directing unconventional computations break this boundary leading to an open algorithmic multiverse – world of unrestricted creativity.

The paper is organized as follows. First, we summarize how the *closed algorithmic universe* was created and what are advantages and disadvantages of living inside such a closed universe. Next, we describe the breakthrough brought about by the creation of super-recursive algorithms. In Section 4, we analyze super-recursive algorithms as cognitive tools. The main effect is the immense growth of cognitive possibilities and computational power that enables corresponding growth of information processing devices.

**The Closed Universe of Turing Machines and other Recursive Algorithms**

Historically, after having an extensive experience of problem solving, mathematicians understood that problem solutions were based on various algorithms. Construction algorithms and deduction algorithms have been the main tools of mathematical research. When they repeatedly encountered problems they were not able to solve, mathematicians, and especially experts in mathematical logic, came to the conclusion that it was necessary to develop a rigorous mathematical concept of algorithm and to prove that some problems are indeed unsolvable. Consequently, a diversity of exact mathematical models of algorithm as a general concept was proposed. The first models were λ-*calculus* developed by Church in 1931 – 1933, *general recursive functions* introduced by Gödel in 1934, ordinary *Turing machines* constructed by Turing in 1936 and in a less explicit form by Post in 1936, and *partial recursive functions* built by Kleene in 1936. Creating λ-calculus, Church was developing a logical theory of functions and suggested a formalization of the notion of com-putability by means of λ-definability. In 1936, Kleene demonstrated that λ-definability is computationally equivalent to general recursive functions. In 1937, Turing showed that λ-definability is computationally equivalent to Turing machines. Church was so impressed by these results that he suggested what was later called the Church-Turing thesis. Turing formulated a similar conjecture in the Ph.D. thesis that he wrote under Church's supervision.

It is interesting to know that the theory of Frege [1] actually contains λ-calculus. So, there were chances to develop a theory of algorithms and computability in the $19^{th}$ century. However, at that time, the mathematical community did not feel a need of such a theory and probably, would not accept it if somebody created it.

The Church-Turing thesis explicitly mark out a rigid boundary for the algorithmic universe, making this universe closed by Turing machines. Any algorithm from this universe was inside that boundary.

After the first breakthrough, other mathematical models of algorithms were suggested. They include a variety of Turing machines: *multihead, multitape Turing machines, Turing machines with n-dimensional tapes, nondeterministic, probabilistic, alternating* and *reflexive Turing machines, Turing machines with oracles, Las Vegas Turing machines*, etc.; *neural networks* of various types – *fixed-weights, unsupervised, supervised, feedforward,* and *recurrent neural networks*; *von Neumann automata* and general *cellular automata*; *Kolmogorov algorithms finite automata* of different forms – *automata without memory, autonomous automata, automata without output or accepting automata, deterministic, nondeterministic, probabilistic automata*, etc.; *Minsky machines*;

*Storage Modification Machines* or simply, *Shönhage machines*; *Random Access Machines* (RAM) and their modifications - *Random Access Machines with the Stored Program* (RASP), *Parallel Random Access Machines* (PRAM); *Petri nets* of various types – *ordinary* and ordinary *with restrictions*, *regular, free, colored*, and *self-modifying Petri nets*, etc.; *vector machines*; *array machines*; *multidimensional structured model of computation and computing systems*; *systolic arrays*; *hardware modification machines*; *Post productions*; *normal Markov algorithms*; *formal grammars* of many forms – *regular, context-free, context-sensitive, phrase-structure*, etc.; and so on. As a result, the theory of algorithms, automata and computation has become one of the foundations of computer science.

In spite of all differences between and diversity of algorithms, there is a unity in the system of algorithms. While new models of algorithm appeared, it was proved that no one of them could compute more functions than the simplest Turing machine with a one-dimensional tape. All this give more and more evidence to validity of the Church-Turing Thesis.

Even more, all attempts to find mathematical models of algorithms that were stronger than Turing machines were fruitless. Equivalence with Turing machines has been proved for many models of algorithms. That is why the majority of mathematicians and computer scientists have believed that the Church-Turing Thesis was true. Many logicians assume that the Thesis is an axiom that does not need any proof. Few believe that it is possible to prove this Thesis utilizing some evident axioms. More accurate researchers consider this conjecture as a law of the theory of algorithms, which is similar to the laws of nature that might be sup-ported by more and more evidence or refuted by a counter-example but cannot be proved.

Besides, the Church-Turing Thesis is extensively utilized in the theory of algorithms, as well as in the methodological context of computer science. It has become almost an axiom. Some researchers even consider this Thesis as a unique absolute law of computer science.

Thus, we can see that the initial aim of mathematicians was to build a closed algorithmic universe, in which a universal model of algorithm provided a firm foundation and as it was found later, a rigid boundary for this universe supposed to contain all of mathematics.

It is possible to see the following advantages and disadvantages of the closed algorithmic universe.

*Advantages*:

1. Turing machines and partial recursive functions are feasible mathematical models.

2. These and other recursive models of algorithms provide an efficient possibility to apply mathematical techniques.

3. The closed algorithmic universe allowed mathematicians to build beautiful theories of Turing machines, partial recursive functions and some other recursive and subrecursive algorithms.

4. The closed algorithmic universe provides sufficiently exact boundaries for knowing what is possible to achieve with algorithms and what is impossible.

5. The closed algorithmic universe provides a common formal language for researchers.

6. For computer science and its applications, the closed algorithmic universe provides a diversity of mathematical models with the same computing power.

*Disadvantages*:

1. The main disadvantage of this universe is that its main principle - the Church-Turing Thesis - is not true.

2. The closed algorithmic universe restricts applications and in particular, mathematical models of cognition.

3. The closed algorithmic universe does not correctly reflect the existing computing practice.

## The Open World of Super-Recursive Algorithms and Algorithmic Constellations

Contrary to the general opinion, some researchers expressed their concern for the Church-Turing Thesis. As Nelson writes [2], "*Although Church-Turing Thesis has been central to the theory of effective decidability for fifty years, the question of its epistemological status is still an open one.*" There were also researchers who directly suggested arguments against validity of the Church-Turing Thesis. For instance, Kalmar [3] raised intuitionistic objections, while Lucas and Benacerraf discussed objections to mechanism based on theorems of Gödel that indirectly threaten the Church-Turing Thesis. In 1972, Gödel's observation entitled "A philosophical error in Turing's work" was published where he declared that: "*Turing in his 1937, p. 250 (1965, p. 136), gives an argument which is supposed to show that mental procedures cannot go beyond mechanical procedures. However, this argument is inconclu-* *sive. What Turing disregards completely is the fact that mind, in its use, is not static, but constantly developing, i.e., that we understand abstract terms more and more precisely as we go on using them, and that more and more abstract terms enter the sphere of our understanding. There may exist systematic methods of actualizing this development, which could form part of the procedure. Therefore, although at each stage the number and precision of the abstract terms at our disposal may be finite, both (and, therefore, also Turing's number of distinguishable states of mind) may converge toward infinity in the course of the application of the procedure."* [4]

Thus, pointing that Turing disregarded completely the fact that mind, in its use, is not static, but constantly developing, Gödel predicted necessity for super-recursive algorithms that realize inductive and topological computations [5]. Recently, Sloman [6] explained why recursive models of algorithms, such as Turing machines, are irrelevant for artificial intelligence.

Even if we abandon theoretical considerations and ask the practical question whether recursive algorithms provide an adequate model of modern computers, we will find that people do not see correctly how computers are functioning. An analysis demonstrates that while recursive algorithms gave a correct theoretical representation for computers at the beginning of the "computer era", super-recursive algorithms are more adequate for modern computers. Indeed, at the beginning, when computers appeared and were utilized for some time, it was necessary to print out data produced by computer to get a result. After printing, the computer stopped functioning or began to solve another problem. Now people are working with displays and computers produce their results mostly on the screen of a monitor. These results on the screen exist there only if the computer functions. If this computer halts, then the result on its screen disappears. This is opposite to the basic condition on ordinary (recursive) algorithms that implies halting for giving a result.

Such big networks as Internet give another important example of a situation in which conventional algorithms are not adequate. Algorithms embodied in a multiplicity of different programs organize network functions. It is generally assumed that any computer program is a conventional, that is, recursive algorithm. However, a recursive algorithm has to stop to give a result, but if a network shuts down, then something is wrong and it gives no results. Consequently, recursive algorithms turn out to be too weak for the network representation, modeling and study.

Even more, no computer works without an operating system. Any operating system is a program and any computer program is an algorithm according to the general understanding. While a recursive algorithm has to halt to give a result, we cannot say that a result of functioning of operating system is obtained when computer stops functioning. To the contrary, when the operating system does not work, it does not give an expected result.

Looking at the history of unconventional computations and super-recursive algorithms we see that Turing was the first who went beyond the "Turing" computation that is bounded by the Church-Turing Thesis. In his 1938 doctoral dissertation, Turing introduced the concept of a *Turing machine with an oracle*. This, work was subsequently published in 1939. Another approach that went beyond the Turing-Church Thesis was developed by Shannon [7], who introduced the *differential analyzer*, a device that was able to perform continuous operations with real numbers such as operation of differentiation. However, mathematical community did not accept operations with real numbers as tractable because irrational numbers do not have finite numerical representations.

In 1957, Grzegorczyk introduced a number of equivalent definitions of computable real functions. Three of Grzegorczyk's constructions have been extended and elaborated independently to super-recursive methodologies: the *domain approach* [8,9], *type 2 theory of effectivity* or *type 2 recursion theory* [10,11], and the *polynomial approximation approach* [12]. In 1963, Scarpellini introduced the class $\mathbf{M}_1$ of functions that are built with the help of five operations. The first three are elementary: substitutions, sums and products of functions. The two remaining operations are performed with real numbers: integration over finite intervals and taking solutions of Fredholm integral equations of the second kind.

Yet another type of super-recursive algorithms was introduced in 1965 by Gold and Putnam, who brought in concepts of *limiting recursive function* and *limiting partial recursive function*. In 1967, Gold produced a new version of limiting recursion, also called *inductive inference*, and applied it to problems of learning. Now inductive inference is a fruitful direction in machine learning and artificial intelligence.

One more direction in the theory of super-recursive algorithms emerged in 1967 when Zadeh introduced *fuzzy algorithms*. It is interesting that limiting recursive function and limiting partial recursive function were not considered as valid models of algorithms even by their authors. A proof that fuzzy algorithms are more powerful than Turing machines was obtained much later (Wiedermann, 2004). Thus, in spite of the existence of super-recursive algorithms, researchers continued to believe in the Church-Turing Thesis as an absolute law of computer science.

After the first types of super-recursive models had been studied, a lot of other super-recursive algorithmic models have been created: *inductive Turing machines*, *limit Turing machines*, *infinite time Turing machines*, *general Turing machines*, *accelerating Turing machines*, *type* 2 *Turing machines*, *mathematical machines*, δ-Q-*machines*, *general dynamical systems*, *hybrid systems*, *finite dimensional machines* over real numbers, **R**-*recursive functions* and so on.

To organize the diverse variety of algorithmic models, we introduce the concept of an algorithmic constellation. Namely, an *algorithmic constellation* is a system of algorithmic models that have the same type. Some algorithmic constellations are disjoint, while other algorithmic constellations intersect. There are algorithmic constellations that are parts of other algorithmic constellations.

Below some of algorithmic constellations are described.

The *sequential algorithmic constellation* consists of models of sequential algorithms. This constellation includes such models as deterministic finite automata, deterministic pushdown automata with one stack, evolutionary finite automata, Turing machines with one head and one tape, Post productions, partial recursive functions, normal Markov algorithms, formal grammars, inductive Turing machines with one head and one tape, limit Turing machines with one head and one tape, reflexive Turing machines with one head and one tape, infinite time Turing machines, general Turing machines with one head and one tape, evolutionary Turing machines with one head and one tape, accelerating Turing machines with one head and one tape, type 2 Turing machines with one head and one tape, Turing machines with oracles.

The *concurrent algorithmic constellation* consists of models of concurrent algorithms. This constellation includes such models as Petri nets, artificial neural networks, nondeterministic Turing machines, probabilistic Turing machines, alternating Turing machines, Communicating Sequential Processes (CSP) of Hoare, Actor model, Calculus of Communicating Systems (CCS) of Milner, Kahn process networks, dataflow process networks, discrete event simulators, View-Centric Reasoning (VCR) model of Smith, event-signal-process (ESP) model of Lee and Sangiovanni-Vincentelli, extended view-centric reasoning (EVCR) model of Burgin and Smith, labeled transition systems, Algebra of Communicating Processes (ACP) of Bergstra and Klop, event-action-process (EAP) model of Burgin and Smith, synchronization trees, and grid automata.

The *parallel algorithmic constellation* consists of models of parallel algorithms and is a part of the concurrent algorithmic constellation. This constellation includes such models as pushdown automata with several stacks, Turing machines with several heads and one or several tapes, Parallel Random Access Machines, Kolmogorov algorithms, formal grammars with prohibition, inductive Turing machines with several heads and one or several tapes, limit Turing machines with several heads and one or several tapes, reflexive Turing machines with several heads and one or several tapes, general Turing machines with several heads and one or several tapes, accelerating Turing machines with several heads and one or several tapes, type 2 Turing machines with several heads and one or several tapes.

The *discrete algorithmic constellation* consists of models of algorithms that work with discrete data, such as words of formal language. This constellation includes such models as finite automata, Turing machines, partial recursive functions, formal grammars, inductive Turing machines and Turing machines with oracles.

The *topological algorithmic constellation* consists of models of algorithms that work with data that belong to a topological space, such as real numbers. This constellation includes such models as the differential analyzer of Shannon, limit Turing machines, finite dimensional and general machines of Blum, Shub, and Smale, fixed point models, topological algorithms, neural networks with real number parameters.

Although several models of super-recursive algorithms already existed in 1980s, the first publication where it was explicitly stated and proved that there are algorithms more powerful than Turing machines was [13]. In this work, among others, relations between Gödel's incompleteness results and super-recursive algorithms were discussed.

Super-recursive algorithms have different computing and accepting power. The closest to conventional algorithms are inductive Turing machines of the first order because they work with constructive objects, all steps of their computation are the same as the steps of conventional Turing machines and the result is obtained in a finite time. In spite of these similarities, inductive Turing machines of the first order can compute much more than conventional Turing machines [14, 5].

Inductive Turing machines of the first order form only the lowest level of super-recursive algorithms. There are infinitely more levels and as a result, the algorithmic universe grows into the algorithmic multiverse becoming open and amenable. Taking into consideration algorithmic schemas, which go beyond super-recursive algorithms, we come to an open world of information processing, which includes the algorithmic multiverse with its algorithmic constellations. Openness of this world has many implications for human cognition in general and mathematical cognition in particular. For instance, it is possible to demonstrate that not only computers but also the brain can work not only in the recursive mode but also in the inductive mode, which is essentially more powerful and efficient. Some of the examples are considered in the next section.

**Absolute Prohibition in The Closed Universe
and Infinite Opportunities in The Open World**

To provide sound and secure foundations for mathematics, David Hilbert proposed an ambitious and wide-ranging program in the philosophy and foundations of mathematics. His approach formulated in 1921 stipulated two stages. At first, it was necessary to formalize classical mathematics as an axiomatic system. Then, using only restricted, "finitary" means, it was necessary to give proofs of the consistency of this axiomatic system.

Achieving a definite progress in this direction, Hilbert became very optimistic. As a response to the Latin dictum: "*Ignoramus et ignorabimus*" or "*We do not know, we cannot know*", in his speech in Königsberg in 1930, he made a famous statement:

*Wir müssen wissen. Wir werden wissen.*
*(We must know. We will know.)*

Next year the Gödel undecidability theorems were published [15]. They undermined Hilbert's statement and his whole program. Indeed, the first Gödel undecidability theorem states that it is impossible to validate truth for all true statements about objects in an axiomatic theory that includes formal arithmetic. This is a consequence of the fact that it is impossible to build all sets from the arithmetical hierarchy by Turing machines. In such a way, the closed Algorithmic Universe imposed restriction on the mathematical exploration. Indeed, rigorous mathematical proofs are done in formal mathematical systems. As it is demonstrated (cf., for example, [16]), such systems are equivalent to Turing machines as they are built by means of Post productions. Thus, as Turing machines can model proofs in formal systems, it is possible to assume that proofs are performed by Turing machines.

The second Gödel undecidability theorem states that for an effectively generated consistent axiomatic theory $T$ that includes formal arithmetic and has means for formal deduction, it is impossible to prove consistency of $T$ using these means.

From the very beginning, Gödel undecidability theorems have been comprehended as absolute restrictions for scientific cognition. That is why Gödel undecidability theorems were so discouraging that many mathematicians consciously or unconsciously disregarded them. For instance, the influential group of mostly French mathematicians who wrote under the name Bourbaki completely ignored results of Gödel [17].

However, later researchers came to the conclusion that these theorems have such drastic implications only for formalized cognition based on rigorous mathematical tools. For instance, in the 1964 postscript, Gödel wrote that undecidability theorems "do not establish any bounds for the powers of human reason, but rather for the potentialities of pure formalism in mathematics."

Discovery of super-recursive algorithms and acquisition of the knowledge of their abilities drastically changed understanding of the Gödel's results. Being a consequence of the closed nature of the closed algorithmic universe, these undecidability results lose their fatality in the open algorithmic universe. They become relativistic being dependent on the tools used for cognition. For instance, the first undecidability theorem is equivalent to the statement that it is impossible to compute by Turing machines or other recursive algorithms all levels of the Arithmetical Hierarchy [18]. However, as it is demonstrated in [19], there is a hierarchy of inductive Turing machines so that all levels of the Arithmetical Hierarchy are computable and even decidable by these inductive Turing machines. Complete proofs of these results were published only in 2003 due to the active opposition of the proponents of the Church-Turing Thesis [14]. In spite of the fast development of computer technology and computer science, the research community in these areas is rather conservative although more and more researchers understand that the Church-Turing Thesis is not correct.

The possibility to use inductive proofs makes the Gödel's results relative to the means used for proving mathematical statements because decidability of the Arithmetical Hierarchy implies decidability of the formal arithmetic. For instance, the first Gödel undecidability theorem is true when recursive algorithms are used for proofs but it becomes false when inductive algorithms, such as inductive Turing machines, are uti-lized. History of mathematics also gives supportive evidence for this conclusion. For instance, in 1936 by Gentzen, who in contrast to the second Gödel undecidability theorem, proved consistency of the formal arithmetic using ordinal induction.

The hierarchy of inductive Turing machines also explains why the brain of people is more powerful than Turing machines, supporting the conjecture of Roger Penrose [20]. Besides, this hierarchy allows researchers to eliminate restrictions of recursive models of algorithms in artificial intelligence described by Sloman [6].

It is important to remark that limit Turing machines and other topological algorithms [21] open even broader perspectives for information processing technology and artificial intelligence than inductive Turing machines.

**The Open World of Knowledge and The Internet**

The *open world*, or more exactly, the *open world of knowledge*, is an important concept for the knowledge society and its knowledge economy. According to Rossini [12], it emerges from a world of pre-Internet political systems, but it has come to encompass an entire worldview based on the transformative potential of open, shared, and connected technological systems. The idea of an open world synthesizes much of the social and political discourse around modern education and scientific endeavor and is at the core of the *Open Access* (OA) and *Open Educational Resources* (OER) movements. While the term *open society* comes from international relations, where it was developed to describe the transition from political oppression into a more democratic society, it is now being appropriated into a broader concept of an open world connected via technology [22]. The idea of openness in access to knowledge and education is a reaction to the potential afforded by the global networks, but is inspired by the sociopolitical concept of the open society.

*Open Access* (OA) is a knowledge-distribution model by which scholarly, peer-reviewed journal articles and other scientific publications are made freely available to anyone, anywhere over the Internet. It is the foundation for the open world of scientific knowledge, and thus, a principal component of the open world of knowledge as a whole. In the era of print, open access was economically and physically impossible. Indeed, the lack of physical access implied the lack of knowledge access - if one did not have physical access to a well-stocked library, knowledge access was impossible. The Internet has changed all of that, and OA is a movement that recognizes the full potential of an open world metaphor for the network.

In OA, the old tradition of publishing for the sake of inquiry, knowledge, and peer acclaim and the new technology of the Internet have converged to make possible an unprecedented public good: "the world-wide electronic distribution of the peer-reviewed journal literature" [23].

The open world of knowledge is based on the Internet, while the Internet is based on computations that go beyond Turing machines. One of the basic principles of the Internet is that it is always on, always available. Without these features, the Internet cannot provide the necessary support for the open world of knowledge because ubiquitous availability of knowledge resources demands non-stopping work of the Internet. At the same time, classical models of algorithms, such as Turing machines, stop after giving that result. This contradicts the main principles of the Internet. In contrast to classical models of computation, as it is demonstrated in [5], if an automatic system, e.g., a computer or computer network, works without halting, gives results in this mode and can simulate any operation of a universal Turing machine, then this automatic (computer) system is more powerful than any Turing machine. This means that this automatic (computer) system, in particular, the Internet, performs unconventional computations and is controlled by super-recursive algorithms. As it is explained in [5], attempts to reduce some of these systems, e.g., the Internet, to the recursive mode, which allows modeling by Turing machines, make these systems irrelevant.

**Conclusions**

This paper shows how the universe (the world) of algorithms became open with the discovery of super-recursive algorithms, providing more powerful tools for computational cognition and artificial intelligence.

Here we considered only some of the consequences of the open world environment of unconventional algorithms and algorithmic constellations for mathematical (computation-theoretical) cognition. It would be interesting to study other consequences of current break through into an open world of unconventional algorithms and computation.

It is known that not all quantum mechanical events are Turing-computable. So, it would be interesting to find a class of super-recursive algorithms that compute all such events or to prove that such a class does not exist.

It might be methodologically and philosophically interesting to contemplate relations between the Open World of Algorithmic Constellations and the Open Science in the sense of Nielsen [24]. For instance, one of the pivotal features of the Open Science is accessibility of research results on the Internet. At the same time, as it is demonstrated in [5], the Internet and other big networks of computers are always working in the inductive mode or some other super-recursive mode. Moreover, actual accessibility depends on such modes of functioning.

One more interesting problem is to explore relations between the Open World of Algorithmic Constellations with the theoretical framework of Info-computationalism, a synthesis of Pancomputationalism (Naturalist Computationalism) with Informational Structural Realism – the model of a universe as a network of computational processes on informational structures. Info-computationalism connects algorithms with interactive computing in natural (physical) systems [25,26][28]. Connecting new unconventional models of super-recursive algorithms and Algorithmic Constellations with unconventional computations performed by natural systems opens new possibilities for the development of innovative models of physical computation with "Trans-Turing" algorithms and "Non-Von" computing architectures. [27].

**Acknowledgements**

The authors would like to thank Andree Ehresmann, Hector Zenil and Marcin Schroeder for useful and constructive comments on the previous version of this work.

28 Dodig-Crnkovic G., Significance of Models of Computation from Turing Model to Natural Computation. Minds and Machines, ( R. Turner and A. Eden guest eds.) Volume 21, Issue 2 (2011), Page 301.

All the links accessed at 08 06 2012